\documentclass{article}
\title{Non Super-Cell SuperConductivity Of High Tc  Materials }
\author{Fredrick Michael*}
\begin{document}
\maketitle
\begin{abstract}
Recently we have described materials interface transport coupling rigorously utilizing NEGF nonequilibrium Green's functions, and have discussed the Hamiltonian terms that from Green's theorem and boundary conditions can be  rewritten as Self Energy.

We derive the application of our theory to the high $Tc$ Superconducting materials interfaces that are the composition of the high temperature superconducting materials. The derivation models a non super-cell geometry of plaquettes that will describe the superconducting 2D material in abrupt coupling with the material of insulating or normal conducting composition.
\end{abstract}
\section{Derivation}

The superconducting high $Tc$ temperature materials that we will describe are the well known high temperature copper oxides with materials such as Yttrium, Barium, and so on. an example is YtBaCuO${_x}$  . We describe the conductor Copper oxide by an effective electron mass Hamiltonian first for the Copper
\begin{equation}
H_Cu = \sum\limits_{\vec k}^{inf} \epsilon_{\vec k} C_{\vec k} {C_{\vec k}}^{\dagger} + \sum\limits_{\vec k,\vec l, \vec p, \vec q}^{inf} {{C_{\vec k}}^\dagger} {{C_{\vec l}}^\dagger} C_{\vec p} C_{\vec q} 
\end{equation}

The operators utilized with many-particle state functions, the Green's function is derived as
\begin{equation}
{_{Cu}g}^r (\vec k,\omega)=\frac{-i}{\hbar \omega - \epsilon_o (\vec k) -{\Sigma_o} (\vec k, \omega) +i\eta}
\end{equation}
Here the kinetic energy is $\epsilon(\vec k)=\frac{\hbar^2 {\vec k}^2}{2 m_e}$ and when we make the approximation of the effective electron mass $m^*$ which here  is taken to be a constant, the Green's function with $ \epsilon(\vec k)=\frac{\hbar^2 {\vec k}^2}{2 m^*}$ becomes
\begin{equation}
{_{Cu}g}^r (\vec k,\omega)=\frac{-i}{\hbar \omega - \epsilon(\vec k)  +i\eta} \label{eqn3}
\end{equation}

The conducting 2D ribbon is the potentially superconducting material. This is modeled by the simple(!) BCS Bardeen-Cooper-Schrieffer Hamiltonian and the Hamiltonian and the green's function becomes a matrix

\begin{eqnarray}
 {_{Cu}  g}^r =\left( \:\:\:\: \matrix{ {{  g_{nn}}^r}  \:\: {{  g_{ns}}^r} \nonumber \\  {{  g_{sn}}^r}  \:\: {{  g_{ss}}^r}  }  \right) \;\;\;\;\;\;\;\;  \label{eqn4}
\end{eqnarray}

The ribbon of Copper is not a simple metal , it is an alloy in high Tc materials. Therefore we have to modify these equations as to include the alloying material here Oxygen in some ratio. The Hamiltonian now is written as 
\begin{eqnarray}
H_{CuO_x} = \sum\limits_{\vec k}^{inf} \epsilon_{\vec k} C_{\vec k} {C_{\vec k}}^{\dagger} + \sum\limits_{\vec k}^{inf} \epsilon_{\vec k'} C_{\vec k'} {C_{\vec k'}}^{\dagger} + \sum\limits_{\vec k,\vec l, \vec p, \vec q}^{inf} {{C_{\vec k}}^\dagger} {{C_{\vec l}}^\dagger} C_{\vec p} C_{\vec q}  \\  \nonumber
+ \sum\limits_{\vec k',\vec l', \vec p', \vec q'}^{inf} {{C_{\vec k'}}^\dagger} {{C_{\vec l'}}^\dagger} C_{\vec p'} C_{\vec q'}  
+  \sum\limits_{\vec s,\vec t, \vec u, \vec v}^{inf} {{C_{\vec s}}^\dagger} {{C_{\vec t}}^\dagger} C_{\vec u} C_{\vec v}  \\ \label{eqn55}
\end{eqnarray}

The lattice geometry is included in our Hamiltonian by the second quantization derivation, the operators in the position representation acting on the wave functions which include the lattice geometry's symmetry. The interactions here are the $Cu-Cu$ interactions, the $O-O$ interactions, and the $Cu-O$ interactions.
 The Green's function Eq.(\ref{eqn4}) then is the matrix of the normal alloy, and the superconducting alloy  $ {_{(Cu-O)}  g}^r $. The alloy is a perturbed BCS superconductor in a way of considering this model. The perturbation is a coherence building perturbation presumably, as the effective attraction between electrons due to coupling with the lattice is increased by this alloying, and is an insulating mechanism against thermal decohering effects thereby increasing the $Tc$ critical temperature of transition from the normal to the superconducting state. A similar binary alloy mechanism is presumably the cause behind the high $Tc$ of $MgB_2$ .

Simplifications to the Hamiltonian Eq.(\ref{eqn55}) can be made. The Copper Hamiltonian can be approximated again as an effective mass Hamiltonian though 'perturbed'. The Oxygen Hamiltonian also can be approximated as a perturbed effective mass Hamiltonian, and since we are interested in the Copper conductor, as the perturbation to the the electronic bands of the Copper lattice. a further simplification is to approximate the perturbed Copper Hamiltonian effective mass $m^* = \frac{\hbar^2 \vec k}{ \frac{\omega(\vec k)}{\partial \vec k}}=\frac{m_e}{[1 + \frac{\partial \Sigma_{Cu-Cu}}{\partial \vec k} +  \frac{\partial \Sigma_{O-O}}{\partial \vec k} + \frac{\partial \Sigma_{Cu-O}}{\partial \vec k} ]}$ . A simplification is that the alloying does not destroy the formation of bands, the interplay of the attraction between negatively charged electrons and the nearby positively charged nuclei, or the screening of the Coulomb $e^- - e^-$ interaction effect that occurs by the background positive nuclei charge...it merely perturbs it , perhaps a more accurate description would be to say modifies it as the alloying is a large scale effect. An approximation is possible then, and we choose a very simplified one of a new effective mass Hamiltonian of the same form as Eq.(\ref{eqn3}), with $\epsilon(\vec k)=\frac{\hbar^2 \vec k^2}{2 m^{*'}}$.

The material that the $CuO_x$ is in contact with is the material Yttrium, Barium, and other less commonly used insulating material. The Hamiltonian description is similar. There is however no band formation, and no effective mass approximation.  The Hamiltonian for Yttrium is $H_{Yt} => \hbar\omega = \epsilon (\vec k) - \Sigma_{Yt} (\vec k, \omega)$ after the second quantization, and where the interaction potentials are described by $\Sigma_{Yt} (\vec k, \omega)$  . The similar description is made for Barium , and with a similar in form Hamiltonian.  Here simplifications are possible such as the hopping Hamiltonian approach, which when 2nd quantized and then transformed to the continuous representation resembles the effective mass Hamiltonian of the conductor metals, however the approximations are based on effective atoms and excess unscreened $per \:\: the \:\: atom$ charge, hopping transport to nearby nearest neighbor and next to n.n. neighbor atoms with constant potential barriers \cite{mahan1}. 

\section{Coupling Materials of the High Tc Superconductor}
The high Tc superconductor can be described by a coupled interface of the materials we have discussed. The YtBaCuO$_x$
geometry is of a 3D Yttrium coupled to the 2D Copper-Oxygen alloy coupled to the 3D Barium . The interaction between these materials we will view as a problem in quantum transport utilizing the NEGF nonequilibrium Green's functions. That is the superconducting electronic device heterostructure we are describing is the Yt-CuO$_x$-Ba which we will couple in subsequent derivation utilizing a theory of continuous self energy of interface coupling we have recently developed \cite{fred1} which unifies the geometric Green's theorem approach of Feuchtwang \cite{feuchtwang1} and the discrete atomistic self energy approach of Datta \cite{datta1}. 

The coupling is in the direction of the material composition. The extent of the high $Tc$ superconductor is orthogonal to the direction of the coupling. This is solved by extending the material thickness in the direction of the coupling, that is the 3D Yttrium is simplified as a thickness $W_{Yt} , W_{Ba}>>W_{Cu-O}$, and the length and heights we describe by the plaquette method, of large enough size to capture the physics of the problem, and where the semi-infinite extent, an approximation to large extent here, is modeled by the periodic boundary conditions we will impose on the solutions of the Green's functions we have discussed in the previous sections.

The coupling via the continuous self energy method we do not discuss in detail, we refer the reader to our previous work \cite{fred1}. The summary is that the Hamiltonians are modified by the self energy due to the coupling in the $\hat z$ direction of Hamiltonian $h(z) =const.( [\frac{\partial}{\partial z},\delta({z- Z_{>} + \nu})]_{+} +  [\frac{\partial}{\partial z},\delta({z- Z_{<} + \nu})]_{+} )$ and where here $Z_> - Z_< = W_{Cu-O_x}$ and $[...]_{+}$ is the additive commutator. The retarded self energy is $\Sigma^r =\frac{\hbar^2}{2m_\alpha}(\gamma_{D\alpha}-{\gamma_\alpha})-\frac{1}{g^r (\alpha,\alpha)}$, the $\gamma= -ik_\alpha$ and $\gamma_{D\alpha}$ chosen as per the partition of the Hamiltonian and according to the boundary conditions chosen. This is for the homogeneous boundary conditions case, and for simplicity we derive Dirichlet boundary conditions $g(\vec x,\vec x')|_{z,z'=Z_{<,>}}=0$ where the self energy becomes $\gamma_\alpha$ with appropriate energy cutoffs.

The quantum transport across the abruptly coupled interface between the Yttrium, the Copper alloy, the Barium is by the derived  nonequilibrium Green's function 
\begin{equation}
G^+ = (1+G^r \Sigma^r) g^+ (1+ G^a \Sigma^a)+ G^r \Sigma^+ G^a
\end{equation}

The self energy is the sum of the coupling $\Sigma = \Sigma_{W_>} + \Sigma_{W_<}$ and where we have written the Yt-CuO$_x$ interface as $W_>$ and the CuO$_x$-Ba here  as $W_{<}$.

A simplified full green's function is to write it in terms of the perturbed by the coupling self energy green's functions $G^r = G^r (\vec x, \vec x')\theta({z-W_>})\theta(z'- W_<) + G^r (\vec x, \vec x')\theta({z+W_>})\theta(z'+ W_>) \theta({z-W_<})\theta(z'- W_<)+  G^r (\vec x, \vec x')\theta({z+W_<})\theta(z'+ W_<)$.

Here the Green's functions are modified Green's functions that aside from interactions include the coupling self energy. For example in the 2D Copper-Oxygen device region, the Hamiltonian is the BCS Hamiltonian and the self energy due to coupling $H=> \hbar\omega =\epsilon(\vec k) - V_{BCS} \Sigma (\vec k, \omega) + \Sigma_{W_<} (\vec k) + \Sigma_{W_>} (\vec k) $.... In the Yttrium region the Hamiltonian is $H_{Yt} =\epsilon(\vec k) + \Sigma_{W_>}$ and so on.

The nonequilibrium Green's function is the matrix equation that is the normal, the superconducting, and the interacting Hamiltonians. As we have rewritten the Hamiltonians to include the coupling self energies, we can rewrite the normal-superconducting Green's function as the quantum transport matrix
\begin{eqnarray}
G^+ =\left(  \:\:\:\: \matrix{   { _{nn}G^+}   \:\:  { _{ns}G^+}  \nonumber \\  { _{sn}{^\dagger}G^+}  \:\: { _{ss}G^+}  }  \right).
\end{eqnarray} 

Other partitionings are possible as we have the freedom to partition the Hamiltonians that comprise the Green's functions. With this partitioning of the Hamiltonians, the superconducting transition is obtained by the Off diagonal long range order ODLRO $\Delta =  { _{ns}G^+} -  { _{ns}{^\dagger}G^+}$, and here $\dagger$ indicates the transpose. The number of particles and the rate of transitions can be obtained from the DLRO diagonal trace and averaging w.r.t. the diagonal trace as $<N>=\int { _{nn}G^+} + \int { _{ss}G^+} $ .  As mentioned one can take the partitioning in a different order or entirely differently, an example is redefining the $G^r => \Delta^r$ and rewriting the NEGF as $G^+ = (1+\Delta^r \Sigma^r){\Delta_o}^+ (1+\Delta^a \Sigma^a)+ \Delta^r \Sigma^+ \Delta^a $.

The temperature of transition, the critical temperature, is evaluated by the integration of the say the retarded Green's function and obtaining the point at which the normal-superconducting Green's function diverges. The point of divergence is where the phase transition occurs and the Temperature is obtained from the dispersion relation evaluation with the distribution and the temperature. Alternatively the thermodynamics of the specific heat can be used to obtain the temperature of the transition.

\section{conclusion}
In this article we have derived a theory of the High temperature superconductors utilizing the nonequilibrium quantum transport theoretical formalism. The resultant equations are relatively simple, and can obtain the temperature of transition by regular phase transition methods, as we have modified the BCS theory minimally by the inclusion of coupling of the 2D conductor alloy to the interface coupled materials.

The temperature of the transition of the superconductor is modified by the coupling of the 3D materials Yttrium, Barium and what else as can be seen by the following discussion. The Total energy is used in the thermodynamics to obtain the specific heat, and for this BCS superconductivity description of the phase transitions, there is a discontinuity at the point of the phase transition from which the critical temperature of the phase transition can be obtained. The energy is the sum integral $<E>=\int \hbar \omega \Delta_{DLRO} d\vec k d\omega$, and as $\hbar\omega= \epsilon(\vec k)+ \Sigma_{(BCS +\Sigma_{W_<} + \Sigma_{W_>})}$ , the energy is shifted shifting the point at which the here BCS description of the superconductivity discontinuous phase transition and its critical temperature occurs.  

This shifting of the critical temperature is also modified by the cohering effects of other perturbations aside from large scale modifications by coupling to 3D materials the 2D conductor alloy. The addition of fields, electric, magnetic, can if applied properly build the coherence of the superconducting state. This coherence building can be by external fields, and/or by materials that are Ferromagnetic, Anti Ferromagnetic,  doped, alloys, and other methods such as geometric arrangements. This derivation for the high $Tc$ superconductors then allows us to pursue these coherence building analytics and we pursue the possibility of designing heterostructure high $Tc$ superconductors in future work.

\pagebreak

\end{document}